\documentclass[usenatbib]{mn2e}

\usepackage{graphicx}
\usepackage{lscape}
\usepackage{longtable}

\title[Discovery of PSR~J2043+1711 at the Nan\c cay Radio Telescope]{Discovery of the millisecond pulsar PSR~J2043+1711 in a \emph{Fermi} source with the Nan\c cay Radio Telescope}

\author[L.~Guillemot et al.]{
L.~Guillemot,$^{1,2}$
P.~C.~C.~Freire,$^{1,3}$
I.~Cognard,$^{4,5}$
T.~J.~Johnson,$^{6}$
Y.~Takahashi,$^{7}$
\newauthor
J.~Kataoka,$^{7}$
G.~Desvignes,$^{1}$
F.~Camilo,$^{8}$
E.~C.~Ferrara,$^{9}$
A.~K.~Harding,$^{9}$
\newauthor
G.~H.~Janssen,$^{10}$
M.~Keith,$^{11}$
M.~Kerr,$^{12}$
M.~Kramer,$^{10,1}$
D.~Parent,$^{13}$
S.~M.~Ransom,$^{14}$
\newauthor
P.~S.~Ray,$^{15}$
P.~M.~Saz~Parkinson,$^{16}$
D.~A.~Smith,$^{17}$
B.~W.~Stappers,$^{10}$
G.~Theureau$^{4}$
\\
$^{1}$Max-Planck-Institut f\"ur Radioastronomie, Auf dem H\"ugel 69, 53121 Bonn, Germany\\
$^{2}$email: guillemo@mpifr-bonn.mpg.de\\
$^{3}$email: pfreire@mpifr-bonn.mpg.de\\
$^{4}$Laboratoire de Physique et Chimie de l'Environnement, LPCE UMR 6115 CNRS, F-45071 Orl\'eans Cedex 02, \\
and Station de radioastronomie de Nan\c{c}ay, Observatoire de Paris, CNRS/INSU, F-18330 Nan\c{c}ay, France\\
$^{5}$email: icognard@cnrs-orleans.fr\\
$^{6}$National Research Council Research Associate, National Academy of Sciences, Washington, DC 20001, \\
resident at Naval Research Laboratory, Washington, DC 20375, USA\\
$^{7}$Research Institute for Science and Engineering, Waseda University, 3-4-1, Okubo, Shinjuku, Tokyo 169-8555, Japan\\
$^{8}$Columbia Astrophysics Laboratory, Columbia University, New York, NY 10027, USA\\
$^{9}$NASA Goddard Space Flight Center, Greenbelt, MD 20771, USA\\
$^{10}$Jodrell Bank Centre for Astrophysics, School of Physics and Astronomy, The University of Manchester, M13 9PL, UK\\
$^{11}$CSIRO Astronomy and Space Science, Australia Telescope National Facility, Epping NSW 1710, Australia\\
$^{12}$W. W. Hansen Experimental Physics Laboratory, Kavli Institute for Particle Astrophysics and Cosmology, \\
Department of Physics and SLAC National Accelerator Laboratory, Stanford University, Stanford, CA 94305, USA\\
$^{13}$Center for Earth Observing and Space Research, College of Science, George Mason University, Fairfax, VA 22030, \\
resident at Naval Research Laboratory, Washington, DC 20375, USA\\
$^{14}$National Radio Astronomy Observatory (NRAO), Charlottesville, VA 22903, USA\\
$^{15}$Space Science Division, Naval Research Laboratory, Washington, DC 20375-5352, USA\\
$^{16}$Santa Cruz Institute for Particle Physics, Department of Physics and Department of Astronomy and Astrophysics, \\
University of California at Santa Cruz, Santa Cruz, CA 95064, USA\\
$^{17}$Universit\'e Bordeaux 1, CNRS/IN2p3, Centre d'\'Etudes Nucl\'eaires de Bordeaux Gradignan, 33175 Gradignan, France\\
}

\date{}

\begin{document}

\maketitle

\begin{abstract}
We report the discovery of the millisecond pulsar PSR~J2043+1711 in 
a search of a \emph{Fermi} Large Area Telescope (LAT) source with 
no known associations, with the Nan\c cay Radio Telescope.
The new pulsar, confirmed with the Green Bank Telescope, 
has a spin period of 2.38 ms, is relatively nearby ($d \la
2$ kpc), and is in a 1.48 day orbit around a low mass companion,
probably a He-type white dwarf. Using an ephemeris based on
Arecibo, Nan\c cay, and Westerbork timing measurements, pulsed
gamma-ray emission was detected in the data recorded by the
\emph{Fermi} LAT. The gamma-ray light curve and spectral properties
are typical of other gamma-ray millisecond pulsars seen with
\emph{Fermi}. X-ray observations of the pulsar with \emph{Suzaku} and 
the \emph{Swift}/XRT yielded no detection. At 1.4 GHz we observe 
strong flux density variations because of interstellar
diffractive scintillation, however a sharp peak can be observed at
this frequency during bright scintillation states. At 327 MHz the
pulsar is detected with a much higher signal-to-noise ratio and its flux density is
far more steady. However, at that frequency the Arecibo instrumentation 
cannot yet fully resolve the pulse profile.
Despite that, our pulse time-of-arrival measurements have a
post-fit residual rms of 2 $\mu$s. This and the expected
stability of this system has made PSR~J2043+1711 one of the first new
\emph{Fermi}-selected millisecond pulsars to be added to pulsar gravitational
wave timing arrays. It has also allowed a significant measurement of
relativistic delays in the times of arrival of the pulses due to the
curvature of space-time near the companion, but not yet with
enough precision to derive useful masses for the pulsar and the
companion. Nevertheless, a mass for the pulsar between 1.7 and 2.0~M$_{\sun}$ 
can be derived if a standard millisecond pulsar formation model is assumed.
In this article we also present a comprehensive summary of 
pulsar searches in \emph{Fermi} LAT sources with the Nan\c cay 
Radio Telescope to date.  
\end{abstract}


\section{Introduction}

Searches for radio pulsars coincident with \emph{Fermi} Large Area Telescope (LAT)
gamma-ray sources with no known associations have been remarkably
successful, with the discovery of more than 30 millisecond pulsars
(MSPs) up to now \citep[e.g.,][]{Ransom2011,Keith2011,Cognard2011}. In
addition, pulsed gamma-ray emission has been observed for about 20
previously-known radio MSPs \citep{Fermi8MSPs,FermiJ0034,FermiJ1823,Guillemot2012}. 
Finally, the LAT has observed gamma-ray emission from globular clusters with properties that 
are consistent with collective emission from populations of MSPs \citep{FermiGCs,Kong2010}. 
MSPs are therefore an important class of gamma-ray sources.

MSPs are rapidly rotating neutron stars, characterized
by small rotational periods ($P \la 30$ ms) and period
derivatives ($\dot P \la 10^{-17}$). 
These pulsars are thought to be ``recycled'', spun-up 
to millisecond periods by the accretion of matter and thus transfer of 
angular momentum from a binary companion \citep{Bisnovatyi1974,Alpar1982}.
More than 80\% of them are in
binary systems, which makes searches for MSPs less sensitive than for
normal pulsars. In addition, the MSPs discovered in \emph{Fermi} LAT 
unassociated sources are widely distributed in Galactic latitude, whereas 
most radio pulsar surveys have concentrated on the Galactic plane. With its 
unprecedented sensitivity and localization accuracy \citep[see][]{FermiLAT}, 
the LAT directs radio telescopes to high latitude unassociated sources that 
could be unknown pulsars, missed by previous radio surveys. Moreover, 
positional uncertainties in the \emph{Fermi} LAT Second Source Catalog 
\citep[2FGL;][]{Fermi2FGL} are comparable to typical radio beam sizes, 
making radio pulsation searches very efficient.

Despite the numerous discoveries, 30\% of the sources in the 2FGL 
catalog remain unassociated and could potentially hide unknown 
gamma-ray pulsars. Pulsars seen by
the \emph{Fermi} LAT are characterized by gamma-ray spectra with sharp
cutoffs at a few GeV and low flux variability
\citep{Fermi8MSPs,FermiPSRCatalog}. Observations of \emph{Fermi} LAT
unassociated sources from the First Source Catalog
\citep[1FGL][]{Fermi1FGL} with curved spectra at the Nan\c cay Radio Telescope (NRT)
have previously yielded the discovery of two radio and gamma-ray MSPs,
PSRs J2017+0603 and J2302+4442 \citep{Cognard2011}. In this article 
we describe a third NRT discovery, PSR~J2043+1711, an MSP in a binary system 
located at the position of a \emph{Fermi} LAT source with pulsar-like properties and no
previously-known, plausible associations.

This paper is organized as follows: in Sections~\ref{discoveryobs} and
\ref{Areciboobs} we describe the radio observations of PSR~J2043+1711.
In Section~\ref{timingobs} we discuss the timing
analysis of this MSP using the radio and the LAT data. In
Sections~ \ref{gammaanalysis} and \ref{xrayanalysis} we describe the
analysis of \emph{Fermi} LAT data for the MSP, observed to emit
pulsed gamma rays, and the results of \emph{Suzaku} observations
yielding no detection of X-ray emission from the pulsar.
In Section~\ref{sec:discussion}, we discuss some of the scientific results,
including the modeling of the observed radio and gamma-ray light curves 
of PSR~J2043+1711 in the context of theoretical models of emission in 
the outer magnetosphere of pulsars, and the detection of the Shapiro delay induced by the companion
star at superior conjunction, which allowed us to place constraints on
the neutron star mass (Section~\ref{sec:masses}) and the detection of the
proper motion and consequent limits on the distance to the
pulsar (Section~\ref{sec:mi}). Furthermore, in Section~\ref{sec:mi}
we use the measured gamma-ray energy density to derive tighter upper
limits for the distance. Once a precise distance is measured, these
can be used to derive a lower limit for the moment of
inertia of the star. We present a final summary and discuss some
scientific prospects in Section~\ref{conclusion}.

\section{Observations and data analysis}

\subsection{Discovery observations}
\label{discoveryobs}

The \emph{Fermi} LAT catalog source 2FGL~J2043.2+1711 was
already listed in the First Source Catalog as 1FGL~J2043.2+1709. It
has no known counterpart, and has spectral and variability properties
that made it a plausible gamma-ray pulsar, with a curvature index of
12.0 and a variability index of 12.2 in the 1FGL catalog \citep[see][for 
definitions of the curvature and variability indices]{Fermi1FGL}. 
A curvature index larger than 11.34 
indicates that the source has a spectrum which departs from a 
pure power-law at the 99\% confidence level, while a variability 
index larger than 23.21 implies that the source shows evidence of flux 
variability at the 99\% confidence level. 
With its lack of known associations, its gamma-ray
properties that are reminiscent of those of known pulsars, and high
Galactic latitude ($b = -15.29\degr$), 1FGL~J2043.2+1709 satisfied
all the criteria used for selecting the sources eligible for the
original search for pulsars in \emph{Fermi} LAT unassociated sources
at Nan\c cay that led to the discovery of PSRs J2017+0603 and
J2302+4442 \citep{Cognard2011}, except for the spatial localization:
the semi-major axis of the 1FGL source 95\% confidence ellipse
($\theta_{95}$) was larger than the conservative cut of 3$\arcmin$ used for
the latter search in order for sources to be well covered by the Nan\c cay 
beam, which has a width at half maximum of 4$\arcmin$ in right ascension 
and 22$\arcmin$ in declination. With additional data, the source localization 
improved ($\theta_{95} = 2.9\arcmin$ in the 2FGL catalog), making 2FGL
J2043.2+1711 an excellent candidate source for radio pulsar searches
with the NRT\footnote{In addition to the successful search strategy just 
described, a number of other NRT observations of other \emph{Fermi} 
LAT sources are documented in the Appendix.}.

A first 1-hr observation of the source was made on 2009
November 21 at the NRT, using the modified Berkeley-Orl\'eans-Nan\c
cay (BON) instrumentation \citep{Cognard2006} at 1.4 GHz, 
with a $512 \times 0.25$ MHz incoherent filter bank sampled every 32 $\mu$s. The data
were dedispersed in $\sim$2000 dispersion measure (DM) values, up to
1244 pc cm$^{-3}$, and processed using acceleration and single pulse
search techniques as provided by the \textsc{Presto} package
\citep{presto}. No pulsations were observed in this initial observation. Similarly, 
the analysis of a second observation taken on 2009 December 2 yielded no 
detection. Nevertheless, a 19$\sigma$ candidate with a rotational period of 
2.379 ms and a DM of 20.7 pc cm$^{-3}$ was observed in the third observation, taken 
on 2009 December 12. Seven subsequent observations of the \emph{Fermi} 
LAT source at 1.4 GHz with the NRT yielded no re-detection, casting doubts 
on the presence of a pulsar in this \emph{Fermi} LAT source. The 2.379 ms candidate was finally confirmed 
with observations made at the GBT telescope at 350 MHz with the GUPPI
backend\footnote{https://safe.nrao.edu/wiki/bin/view/CICADA/NGNPP} 
during a survey of \emph{Fermi} LAT unassociated sources \citep[see][]{Hessels2011},
and at the Arecibo telescope at 327 MHz with the Wide-band Arecibo
Pulsar Processors \citep[WAPPs;][]{Dowd2000}. Substantial
accelerations of the rotational period across the confirmation
observations were measured, indicating that the pulsar is in a binary
system. 

\subsection{Arecibo Observations}
\label{Areciboobs}

\begin{figure}
\includegraphics[scale=0.45]{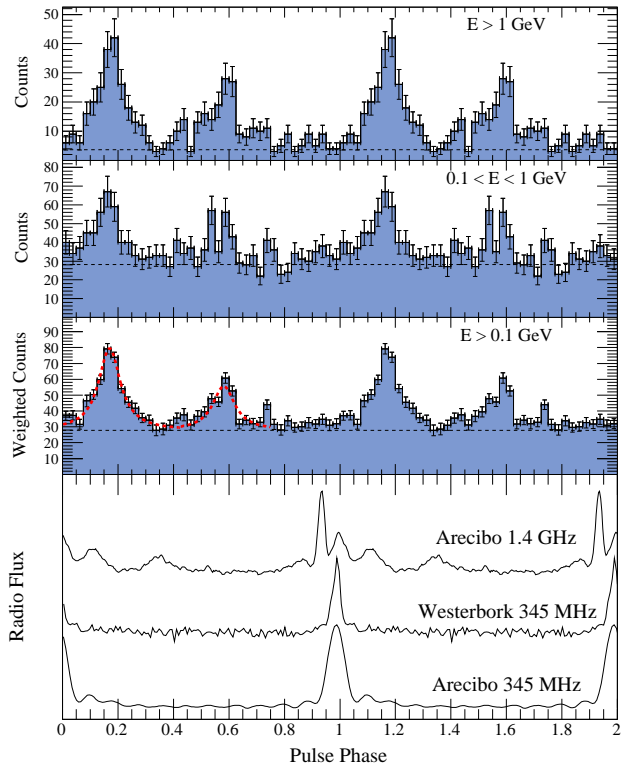}
\caption{Multi-wavelength light curves of PSR~J2043+1711. The bottom 
panel shows radio profiles recorded at the Arecibo telescope at 345 MHz 
(40-min integration), at the Westerbork Synthesis Radio Telescope at 
345 MHz (3.8-hr integration) and at the Arecibo telescope at 1.4 GHz 
(1-hr integration). The third panel shows a 40-bin gamma-ray profile
  obtained by selecting events recorded by the \emph{Fermi} LAT within
  $5\degr$ from the pulsar and with energies above 0.1 GeV, and
  weighting each event by its probability of originating from the
  pulsar. The fitting functions are shown as dashed lines in the first 
  cycle. The top two panels show non-weighted \emph{Fermi} LAT light
  curves for events recorded within $0.8\degr$ of the pulsar, with
  energies between 0.1 and 1 GeV, and above 1 GeV,
  respectively. No gamma-ray features significantly narrower than the 
  bin width used here were observed. Horizontal dashed lines indicate 
  gamma-ray background levels. Two rotations are shown for clarity. \label{lightcurves}}
\end{figure}

Radio light curves of PSR~J2043+1711 recorded with the 
Arecibo telescope are shown in Figure \ref{lightcurves}. At
1.4 GHz the pulsar is observed to exhibit dramatic radio flux
variations, explaining the several unsuccessful attempts to confirm
the pulsar at Nan\c cay following the discovery. Even with Arecibo it
can be a difficult object: during one 30-min observation
the pulsar was not visible in one of the WAPPs, centered at
1.46 GHz and with 50 MHz of bandwidth. Assuming a system temperature
of $T_{sys} = 25$ K, a gain of $G = 10$ K Jy$^{-1}$, a pulse width of 6.25\%
(8 bins above average out of a total of 128), and a signal-to-noise 
ratio (SNR) smaller than 3, this corresponds to an instantaneous upper limit on the radio flux
density of 5.1 $\mu$Jy. However, on most occasions the pulsar is
detectable with Arecibo and the high SNR light
curve was recorded during a bright scintillation state. The 1.4-GHz
light curve is complex, with several pulsed components. A 
total of 4 useful observations of the pulsar at 1.4 GHz have been carried out 
with Arecibo, between 2010 November 20 and 2011 August 24. 
The average radio flux density for these observations was of 
the order of 10 $\mu$Jy. 

At 327 MHz the flux density is far more steady, and for that reason we
carried out the bulk of the Arecibo timing at this frequency. The
pulsar was observed 64 times between 2010 July 17 and 2011 August 25 with
average integrations of 35 minutes using the 327 MHz receiver ($G =
11$ K Jy$^{-1}$, $T_{sys} = 113$ K). For most observations we use the
4 WAPP spectrometers in parallel. Each WAPP makes a
3-level digitization of the analog voltages over a 12.5 MHz band for both
linear polarizations, autocorrelating these for a total of 512
lags. The data are then integrated for $t_s = 64$ $\mu$s and the
orthogonal polarizations added in quadrature are written to disk. Their
bands are centered at 308.25, 320.75, 333.25 and 345.75 MHz, and
together they cover the full 50 MHz band provided by the receiver.
For all Arecibo observations, the lags were Fourier transformed to
generate power spectra. These were dedispersed at the nominal DM of
the pulsar and folded modulo its spin period using the \textsc{Presto}
pulsar software package, generating pulse profiles that are then
stored for later analysis. Ten Arecibo observations at 345 MHz 
were averaged to derive a radio flux density of $S_{345} = 1.2$ mJy 
with a standard deviation of 0.2 mJy, assuming a pulse width of 6.25\%. 

Note that this pulse width is overestimated because of instrumental limitations: 
with the WAPPs the pulses are seen with an effective width at half maximum 
$dt$ given by the sum in quadrature of the intrinsic width at half maximum $dt_i$, 
the sampling time $dt_s$ and the dispersive smearing $dt_d$. At an observing frequency 
$\nu\,=\,$0.345 GHz, we have $dt_i \sim 75-80 \, \mu$s (see below). 
The sampling time is 64 $\mu$s, and with a bandwidth of 12.5 MHz for each WAPP, and 512 spectral channels 
across that bandwidth, the dispersive smearing $dt_d$ \citep[see e.g. Equation (5.2) of][]{Handbook} 
is calculated to be $\sim 120\, \mu$s. Therefore, we have 
an effective width of $\sim 156\, \mu$s, twice as much as the intrinsic pulse width.
The intrinsic pulse profile was determined using 
the PuMa2 backend \citep{Karuppusamy2008} 
at the Westerbork Synthesis Radio Telescope (WSRT), in the
Netherlands, which is capable of coherent dedispersion for a total
bandwidth of 80 MHz centered at 345 MHz (see Figure \ref{lightcurves}).

\subsection{Timing analysis}
\label{timingobs}

The best detections of the pulsar at each instrument and frequency
were used to derive ``standard'' pulse profiles. For the dominant Arecibo 
327 MHz dataset, we made 1 time of arrival (TOA) per WAPP for every 500 s of 
observations on average, by cross-correlating the pulse profiles with the standard profiles 
in the Fourier domain \citep{Taylor1992}. We also extracted 32 TOAs from 
the 1.4 GHz Arecibo observations. This resulted in a total of 1029 Arecibo TOAs.
In addition, 18 NRT TOAs have been recorded at 1.4
and 1.6 GHz from 2010 August 16 to 2011 August 15 
using the procedure and instrumentation described in
\citet{Cognard2011}. Six TOAs were recorded at 345 MHz with the WSRT 
between 2011 March 18 and 2011 August 20. Finally,
gamma-ray pulsations from PSR~J2043+1711 were detected in the
\emph{Fermi} LAT data after the first months of radio timing following
the discovery (see Section \ref{gammaanalysis}), which allowed us to
measure TOAs using the maximum likelihood techniques described in
\citet{Ray2011} and recover phase-coherence across the entire
\emph{Fermi} LAT dataset. A total of 13 gamma-ray TOAs with average 
uncertainty 25.3 $\mu$s and corresponding
to at least 3$\sigma$ detections were extracted between 2008 September 12
and 2011 June 10 by selecting photons with
energies greater than 0.5 GeV and with reconstructed directions found
within $1\degr$ of the pulsar. 

We carried out subsequent TOA analyses using the \textsc{tempo2}
software package \citep{tempo2}. For the conversion of Terrestrial Time (TT) 
TOAs to Coordinated Barycentric Time (TCB) we used the DE/LE~421 
solar system ephemeris \citep{Folkner2008}. The differences between 
observed and predicted barycentric TOAs (the timing residuals) were weighted 
in the fit according to the estimated uncertainty of each TOA. The resulting 
timing parameters and their 1$\sigma$ uncertainties are
presented in Table \ref{ephem}. The orbit of PSR~J2043+1711 has very
low eccentricity, therefore we used the ``ELL1H'' orbital model
\citep{Freire2010,Lange2001} to model it. This yields Keplerian
(semi-major axis of the pulsar orbit projected along the line of sight,
$x$, orbital period $P_b$, epoch of ascending node $T_0$, eccentricity 
$e$ and longitude of periastron, $\omega$) and post-Keplerian 
(orthometric amplitude $h_3$ and orthometric ratio $\varsigma$)
parameters that are weakly correlated with each other.

This ephemeris describes the TOAs well, with a reduced $\chi^2$ of
2.1 for 1044 degrees of freedom. In order to estimate the timing parameters with realistic
uncertainties, we adjusted the uncertainty estimates of each individual timing 
dataset using EFAC parameters so that $\chi^2/n_{\rm free}$ is equal to 1 in every case. The
uncertainties quoted in Table~\ref{ephem} were derived in this fit by
\textsc{tempo2}, except where stated otherwise. The 1029 Arecibo TOAs
have a post-fit rms uncertainty of 2.13 $\mu$s, despite the
aforementioned smearing caused by the instrumentation used to date.
For this reason PSR~J2043+1711 has been added to the International 
Pulsar Timing Array (IPTA) \citep{Hobbs2010}.

There is clearly scope for further improvement in the timing precision
of this object, given the fact that the pulse width measured with the
WAPPs is twice as wide as the intrinsic pulse width. This implies that,
with a broadband coherent dedispersion instrument working at Arecibo,
the measured peak flux density should be about twice as large and the
rise time about half as long, implying about four times the
current timing precision.

Note that the DM value and its uncertainty were measured independently 
from the analysis described here. To measure the DM we built a dataset 
of 308.25, 320.75, 333.25 and 345.75 MHz Arecibo TOAs by cross-correlating 
the individual pulse profiles with a single standard profile, to prevent any 
phase shifts caused by the usage of different template profiles. This dataset 
was then fitted for the DM, yielding the best-fit value and corresponding 
uncertainty listed in Table~\ref{ephem}. 

Because the radio timing data cover only slightly more than a year,
no proper motion could be measured with these data alone. However,
with the gamma-ray TOAs covering  approximately three years, we could
measure a significant proper motion (see Table~\ref{ephem}).
It is clear that the proper motion measurement depends strongly on the gamma-ray 
timing. We therefore checked the proper motion values and their associated uncertainties 
by studying the influence of the LAT timing on the measurement. We first repeated the 
analysis described above using different numbers of gamma-ray TOAs, 
from 5 to 25 with a step of 1. This yielded average proper motion values that are 
consistent with those listed in Table~\ref{ephem}, within standard deviations on 
$\mu_\alpha \cos(\delta)$ and $\mu_\delta$ of 0.1 and 0.4 mas yr$^{-1}$, respectively. Also, 
we made 1000 realizations of a Monte-Carlo simulation in which gamma-ray TOAs 
were generated based on the timing solution given in Table~\ref{ephem}, and were then 
perturbed so that the residuals have the same rms as those of the actual gamma-ray TOAs 
($\sim$ 30 $\mu$s). The uncertainties were finally shuffled from the actual gamma-ray 
TOAs. Again, this yielded average $\mu_\alpha \cos(\delta)$ and $\mu_\delta$ values that are compatible 
with the values reported here, within standard deviations of 0.6 and 0.7 mas yr$^{-1}$, respectively. 
Combining the standard deviations obtained from these two studies, we estimate that the 
gamma-ray TOAs introduce systematic uncertainties on $\mu_\alpha \cos(\delta)$ and $\mu_\delta$ 
of approximately 1 mas yr$^{-1}$.

\begin{table*}
\caption{Measured and derived parameters for PSR~J2043+1711. 
Numbers in parentheses are the nominal 1$\sigma$ 
\textsc{tempo2} uncertainties in the least-significant digits quoted. 
The distance was estimated using the NE2001 model of Galactic free 
electron density \citep{NE2001}. We assumed an uncertainty on this 
distance estimate of 20\%. Using this distance estimate and the measured 
proper motion, $\mu_T$, we calculated the period derivative 
corrected for the Shklovskii effect \citep{Shklovskii1970}, $\dot P_{corr}$, 
and used that value to derive $\dot E$, $B_s$ and $B_{LC}$. 
Note that $\dot E$, $B_s$, $B_{LC}$ and $\eta$ were calculated 
assuming a moment of inertia $I$ of $10^{45}$ g cm$^2$. 
For proper motion parameters and for the total proper motion, 
the first quoted uncertainties are the 1$\sigma$ statistical 
uncertainties from \textsc{tempo2} and the second ones are systematic 
(see Section \ref{timingobs} for details on the calculation of the systematic uncertainties). 
For gamma-ray parameters the first quoted uncertainties are statistical and the second 
are systematic, and correspond to the differences observed when doing the 
spectral analyses with the P6\_V3 IRFs and associated diffuse models.\label{ephem}}
\begin{footnotesize}
\begin{tabular}{lc}
\hline
R.A. (J2000)\dotfill & $20^{\rmn{h}} 43^{\rmn{d}} 20\fs 88309(5)$ \\
Decl. (J2000)\dotfill & $+17\degr 11\arcmin 28\farcs 948(1)$ \\
Rotational period, $P$ (ms)\dotfill & 2.37987896026(4) \\
Apparent period derivative, $\dot P$ ($10^{-21}$)\dotfill & 5.24(2) \\
Proper motion in right ascension, $\mu_{\alpha} \cos(\delta)$ (mas yr$^{-1}$)\dotfill & $- 7 \pm 1 \pm 1$ \\   
Proper motion in declination, $\mu_{\delta}$ (mas yr$^{-1}$)\dotfill & $- 11 \pm 2 \pm 1$ \\
Epoch of ephemeris, $T_0$ (MJD)\dotfill & 55400.00019 \\
Dispersion measure, DM (cm$^{-3}$ pc)\dotfill & 20.70987(3) \\
Orbital period, $P_b$ (d)\dotfill & 1.482290809(2) \\
Projected semi-major axis, $x$ (lt s)\dotfill & 1.6239614(1) \\
Epoch of ascending node, $T_{asc}$ (MJD)\dotfill & 55253.8038503(6) \\
$\eta \equiv e \sin{\omega}$ ($10^{-6}$)\dotfill & $-$2.1(1) \\
$\kappa \equiv e \cos{\omega}$ ($10^{-6}$)\dotfill & $-$2.6(1) \\
$h_3$ ($\mu$s) \dotfill & 0.63(7) \\
$\varsigma$ ($\mu$s) \dotfill & 0.87(5) \\
Span of timing data (MJD)\dotfill & 54729.1 --- 55798.2 \\
Number of TOAs\dotfill & 1066 \\
RMS of TOA residuals ($\mu$s)\dotfill & 2.1 \\
Solar system ephemeris model\dotfill & DE421 \\
Units\dotfill& TCB \\
Flux density at 327 MHz, $S_{327}$ (mJy)\dotfill & $1.2 \pm 0.2$ \\
\hline
\multicolumn{2}{c}{Derived parameters} \\
\hline
Orbital eccentricity, $e$ ($10^{-6}$)\dotfill & 3.4(1) \\
Mass function, $f$ ($M_{\sun}$)\dotfill & 0.00209287(8) \\
Minimum companion mass, $m_c$ ($M_{\sun}$)\dotfill & $\geq$ 0.173 \\
Galactic longitude, $l$ ($\degr$)\dotfill & 61.92 \\
Galactic latitude, $b$ ($\degr$)\dotfill & $-15.31$ \\
Distance inferred from the NE2001 model, $d$ (kpc)\dotfill & $1.8 \pm 0.4$ \\
Total proper motion, $\mu_{T}$ (mas yr$^{-1}$)\dotfill & $13 \pm 2 \pm 1$ \\  
Shklovskii-corrected period derivative, $\dot P_{corr}$ ($10^{-21}$)\dotfill & $3.6 \pm 0.5$ \\
Spin-down luminosity, $\dot E$ ($10^{34}$ erg s$^{-1}$)\dotfill & $1.1 \pm 0.2$ \\
Surface magnetic field strength, $B_s$ ($10^7$ G)\dotfill & $9.3 \pm 0.7$ \\
Magnetic field strength at the light cylinder, $B_{LC}$ ($10^4$ G)\dotfill & $6.3 \pm 0.5$ \\
\hline
\multicolumn{2}{c}{Light curve parameters} \\
\hline
First peak position, $\Phi_1$\dotfill & $0.17 \pm 0.01$ \\
First peak full width at half-maximum, FWHM$_1$\dotfill & $0.10 \pm 0.02$ \\
Second peak position, $\Phi_2$\dotfill & $0.58 \pm 0.01$ \\
Second peak full width at half-maximum, FWHM$_2$\dotfill & $0.10 \pm 0.03$ \\
Radio-to-gamma-ray lag, $\delta$\dotfill & $0.18 \pm 0.01$ \\
Gamma-ray peak separation, $\Delta$\dotfill & $0.41 \pm 0.02$ \\
\hline
\multicolumn{2}{c}{Gamma-ray spectral parameters} \\
\hline
Spectral index, $\Gamma$\dotfill & $1.4 \pm 0.1 \pm 0.4$ \\
Cutoff energy, E$_c$ (GeV)\dotfill & $3.2 \pm 0.6 \pm 1.0$ \\
Photon flux ($> 0.1$ GeV) (10$^{-8}$ cm$^{-2}$ s$^{-1}$)\dotfill & $2.8 \pm 0.3 \pm 0.9$ \\
Energy flux ($> 0.1$ GeV), $F_\gamma$  (10$^{-11}$ erg cm$^{-2}$ s$^{-1}$)\dotfill & $2.8 \pm 0.2 \pm 0.3$ \\
Luminosity, L$_\gamma$ / f$_\Omega$ (10$^{34}$ erg s$^{-1}$)\dotfill & $\left(1.0 \pm 0.1 \pm 0.1 \right) \times \left(d / 1.8\ \rm kpc\right)^2$ \\
Efficiency, $\eta$ / f$_\Omega$\dotfill & $\left(1.0 \pm 0.1 \pm 0.1\right) \times \left(d / 1.8\ \rm kpc\right)^2$ \\
\hline
\end{tabular}
\end{footnotesize}
\end{table*}

\subsection{Gamma-ray analysis}
\label{gammaanalysis}

The analysis of the LAT data was done using the \emph{Fermi} Science
Tools\footnote{http://fermi.gsfc.nasa.gov/ssc/data/analysis/scitools/overview.html}
(STs) v9r23p1. We selected events recorded between 2008 August 4 and
2011 July 21, with reconstructed directions within a $20\degr \times 20\degr$ region 
centered on the pulsar position, energies between 0.1 and 100 GeV, and zenith angles
below $100\degr$. We further selected ``Source'' class events of
the P7\_V6 instrument response functions (IRFs), and rejected times
when the rocking angle of the telescope exceeded $52\degr$ and when
the Earth's limb infringed upon the region of interest (ROI). The
gamma-ray events were phase-folded using the \emph{Fermi} plug-in
distributed with \textsc{Tempo2} \citep{Ray2011} and the ephemeris
given in Table \ref{ephem}.

To measure the spectral properties of the pulsar we fitted sources in
the ROI using a binned maximum likelihood method, using the \emph{pyLikelihood} module
provided with the STs. The spectral parameters of the 54 2FGL catalog sources 
within $20\degr$ of the pulsar were included in
the model. PSR~J2043+1711 and the eight other pulsars in the field of view 
were modeled as exponentially-cutoff power laws of the form:

\begin{eqnarray}
\frac{dN}{dE} = N_0 \left( \frac{E}{\mathrm{1\ GeV}} \right)^{-\Gamma} \exp \left[ - \left( \frac{E}{E_c} \right) \right] .
\end{eqnarray}

\begin{figure}
\includegraphics[scale=0.45]{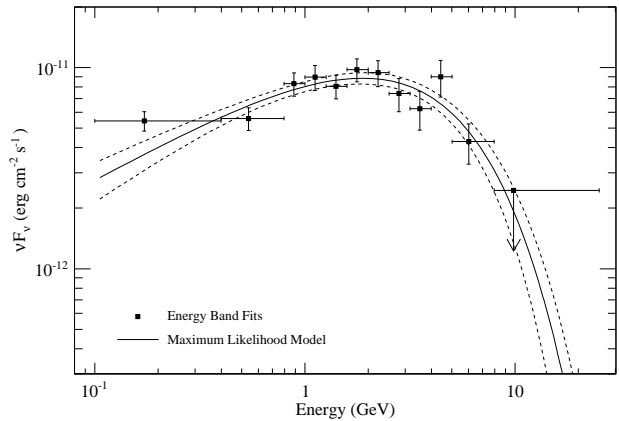}
\caption{Phase-averaged gamma-ray energy spectrum for
  PSR~J2043+1711. The black line shows the best-fit model obtained by
  fitting the \emph{Fermi} LAT data with a simple exponentially cutoff
  power-law form (see Section \ref{gammaanalysis} for details), while 
  dashed error lines indicate 1$\sigma$ errors. Data points are 
  derived from likelihood fits of individual, variable-width energy bands 
  defined by the requirement that the pulsar be detected 
  with a Test Statistic \citep[TS; see][]{Mattox1996} of at least 50. In these 
  bands the pulsar is modeled with a simple power-law form. An 
  upper limit was calculated for the last energy band 
  as the pulsar was not detected with enough significance in that band.\label{spectrum}}
\end{figure}

In this equation, $N_0$ denotes a normalization factor, $\Gamma$ is
the photon index, and $E_c$ is the cutoff energy of the spectrum. The
extragalactic diffuse emission and the residual instrument background
were modeled jointly using the
\emph{iso\_p7v6source} template, while the
Galactic diffuse emission was modeled using the
\emph{gal\_2yearp7v6\_v0} mapcube. 
The spectral parameters of the 12 sources within 
$10\degr$ of PSR~J2043+1711 were left free in the
fit, as were the normalization factors for the diffuse components. 
The nearest source in the catalog, 2FGL~J2031.0+1938, 
is located $\sim 3.7\degr$ away and its gamma-ray energy flux above 
0.1 GeV is $\sim 3$ times smaller than that of PSR~J2043+1711. 
The measurement of spectral properties for the latter object is therefore 
weakly affected by the neighboring sources.
The gamma-ray spectral energy distribution of PSR~J2043+1711 for an
exponentially cutoff power law is shown in Figure
\ref{spectrum}. The best-fit spectral parameters, as well as the
integrated photon and energy fluxes above 0.1 GeV, are
listed in Table \ref{ephem}. 
With the measured energy flux $F_\gamma$, we derived
the gamma-ray luminosity $L_\gamma = 4 \pi f_\Omega F_\gamma d^2$ and
efficiency of conversion of spin-down power into gamma-ray emission,
$\eta = L_\gamma / \dot E$, assuming a geometrical correction factor
$f_\Omega$ (see \citet{Watters2009} for the definition) of 1, which is
typical under outer-magnetospheric gamma-ray emission models
\citep[see e.g.][]{Venter2009}. 
The best-fit spectral parameters for the sources within $10\degr$ were 
compatible within statistical and systematic uncertainties with the values 
published in the 2FGL catalog, with the exception of 2FGL~J2035.4+1058, 
located $6.5\degr$ away from PSR~J2043+1711, for which our flux estimate 
is larger than the 2FGL flux by four standard deviations. The 2FGL source 
J2035.4+1058, which is associated with the blazar PKS 2032+107, 
is nevertheless flagged in the catalog as being variable. Since the time interval 
considered in our analysis was longer than in the 2FGL catalog, 
it is not surprising that the flux we measured differs from the 2FGL one.
We checked the best-fit spectral
parameters for PSR~J2043+1711 by fitting the data with the
\textit{pointlike} likelihood analysis tool \citep{KerrThesis}, and
found results that are consistent with those listed in Table
\ref{ephem} within uncertainties. 

Using the full spectral model obtained with the likelihood analysis
and the tool \emph{gtsrcprob}, we calculated the probabilities that
each gamma-ray event originates from the pulsar. The
probability-weighted gamma-ray light curve of PSR~J2043+1711 above 0.1
GeV and for events found within $5\degr$ of the pulsar is shown in
Figure \ref{lightcurves}. The weighted \emph{H}-test parameter
\citep{Kerr2011weights} is 433.5, corresponding to a pulsation
significance of 18.5$\sigma$. The upper two phase histograms in Figure
\ref{lightcurves} show gamma-ray light curves for events found within
$0.8\degr$ of the pulsar position, in different energy bands. The
background levels in these light curves have been calculated by
summing the probabilities that events have not been produced by the
MSP, as described in \citet{Cognard2011} and \citet{Guillemot2012}. 
We verified that the weighted light curve did not show 
indications for emission features significantly narrower than the 
bin width used, by analyzing the same light curve with five and 
ten times the number of bins, finding no statistically significant 
component other than those seen in Figure \ref{lightcurves}.

We measured the phase $\Phi_i$ and full width at half 
maximum FWHM$_i$ of each gamma-ray
component by fitting the two peaks in the weighted light curve with
Lorentzian functions. The best-fit parameters are listed in Table
\ref{ephem}, along with the radio-to-gamma-ray lag, $\delta = \Phi_1 -
\Phi_r$, where $\Phi_r = 0.99$ is the phase of the maximum of the 345
MHz radio profile, and the separation between the two gamma-ray peaks,
$\Delta = \Phi_2 - \Phi_1$. 

The gamma-ray light curve shape of PSR~J2043+1711 is similar to those
of other gamma-ray MSPs \citep[see
  e.g.][]{Fermi8MSPs,Cognard2011}, suggesting that the gamma-ray
emission is produced at high altitudes in their magnetospheres
\citep{Venter2009}. Likewise, the spectral properties of the MSP are
similar to those of other gamma-ray pulsars
\citep{FermiPSRCatalog}. The large efficiency $\eta \sim 100$\% indicates 
that the distance is very likely overestimated by the NE2001 model. 
In Section~\ref{sec:discussion} we however discuss the possibility to  
use this large efficiency value to constrain the moment of inertia of 
the star, if the actual pulsar distance is close to the DM distance.

\subsection{X-ray analysis}
\label{xrayanalysis}

On 2010 May 3, the \emph{Fermi} LAT unassociated source
1FGL~J2043.2+1709 was observed by the \emph{Suzaku} X-ray
observatory for 18 ks, as part of an unassociated sources observation
campaign. A description of this campaign and the data reduction, as
well as an X-ray image of the field of view around 1FGL~J2043.2+1709
as seen with \emph{Suzaku} can be found in \citet{Takahashi2012}. 

No significant X-ray source can be seen at the position of
PSR~J2043+1711. Assuming a power-law model with a photon index 
of 2, the X-ray upper limit for PSR~J2043+1711 between 0.5
and 8 keV is calculated to be $\sim$ 4.7 $\times$ 10$^{-14}$ erg cm$^{-2}$
s$^{-1}$ (90\% confidence). 

We also checked the available \emph{Swift}/XRT observations of 
PSR~J2043+1711, and found two, totaling $\sim 10.5$ ks of data. 
No X-ray sources were detected within 2$\arcmin$ of the pulsar position, 
thus confirming the \emph{Suzaku} results.

\section{Discussion}
\label{sec:discussion}

\subsection{Light Curve Modeling}
\label{model}

We have fit the radio and gamma-ray light curves of PSR~J2043+1711
to geometric simulations \citep[assuming the vacuum retarded dipole 
magnetic field geometry of][]{Deutsch1955} using a maximum likelihood 
technique.  The gamma-ray light curves were fit with the
two-pole caustic \citep[TPC; ][]{Dyks2003} and outer-gap \citep[OG;
][]{Cheng1986} models.  For our purposes the TPC model is taken to be
a geometric realization of the slot-gap model \citep{Muslimov2004}.
The radio light curves were fit with a hollow-cone beam, core beam,
and cone plus core beam models following \citet{Story2007}.

The gamma-ray light curve (30 bins per rotation) used for the fits was
constructed using all events from the data described in
Section~\ref{gammaanalysis} with reconstructed directions within
0$\fdg8$ of the pulsar radio position.  Additionally, we used either
the 1.4 GHz or 345 MHz radio profile (also in 30 bins) but report only
results from fits with the former as the latter profile is known to be
too wide; however, fits using the 345-MHz profile were used in
estimating systematic biases in our procedure as discussed below.

We have used simulations with the same parameters ($P = 2.5$ ms) as
those in \citet{Cognard2011} except that we have a resolution of 2.5\%
of the polar cap opening angle ($\theta_{\rm PC}\ \sim\ \sqrt{2\pi R_{NS}/P c}$) 
in gap width.  We scanned over the parameter phase space in order 
to find the best-fit for each model as given in Table~\ref{modTab}.

To account for the fact that these models are relatively simple and
the likelihood surfaces can be steep around maxima, implying
unrealistically small uncertainties, we have rescaled the likelihood
differences by $n_{\rm free} / \left[2 *
  (-\ln(\mathcal{L}_{max}))\right]$, where $\mathcal{L}_{max}$ is the
maximum likelihood value and $n_{\rm free}$ is the degrees of freedom for
the given fit.  Assuming the log-likelihood differences follow a
$\chi^{2}$ distribution this results in the best fit corresponding to
a reduced $\chi^{2}$ = 1.  The best-fit uncertainties given in Table~\ref{modTab}
are 68\% confidence level.  The two widths reported are for the accelerating
($w_{acc}$) and emitting ($w_{em}$) gaps, these are the same for the
TPC model.

Our models are relatively simple and, in the case of the radio
profiles, do not contain as many components as are implied by the
data.  Additionally, the geometry used for the magnetic field cannot
be correct as the magnetosphere will, to some extent, be filled with
charges \citep{Goldreich1969} which will distort the gap geometry.
Therefore, we have attempted to estimate systematic biases in the
best-fit geometries reported in Table \ref{modTab}.  To do this we
refit the data while varying the radio uncertainty by a factor of 2,
varying the gamma-ray background estimate by 5\%, using the 1.4 GHz
radio light curve in 60 bins, and performing fits with gamma-ray light
curves corresponding to the energy ranges 0.1 -- 1 GeV and $\geq$ 1 GeV.
None of these changes strongly affected the gap width parameters.
Changes in geometry were typically $\la 8\degr$; however, the
OG best-fit geometry changed by 25 -- 40$\degr$ in one parameter
for both the core and cone beam fits when the radio uncertainty was
doubled and when fitting only the 0.1 -- 1 GeV gamma-ray light
curve.

\begin{figure}
\includegraphics[scale=0.45]{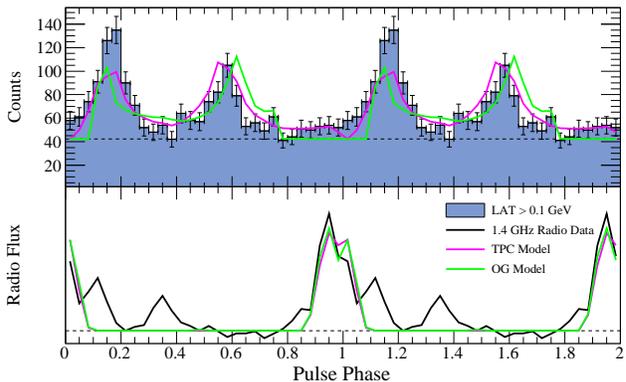}
\caption{Data and best-fit light curves using the gamma-ray data described in Section \ref{model} (top) and the 1.4 GHz radio profile (bottom).  Models corresponding to the TPC fit are shown as dash-dotted lines (solid pink lines online) and for the OG fit as dashed lines (solid green lines online). \label{modLCs}}
\end{figure}

The data and best-fit model light curves are shown in
Figure~\ref{modLCs} using the 1.4 GHz radio profile and the
hollow-cone beam model.  
While the likelihood formally favors the TPC model, both gamma-ray models reproduce 
the qualitative features of the observed gamma-ray light curve (gamma-ray 
peak separation and bridge emission) well, and the radio model is not optimal. 
The OG model therefore cannot be ruled out.
Fits with both the TPC and OG model do
find solutions with $\zeta$ near $80\degr$ which is to be expected
if the spin and orbital axes have (nearly) aligned over time.

It should be noted that the likelihood favors the core beam model in all cases. 
Nevertheless, under these fits we obtain impact parameters ($\beta\ \equiv\ \zeta-\alpha$) 
of order $20\degr$ which  would imply a very faint radio flux for a beam falling off as a 
Gaussian away from the magnetic axis. Such solutions thus seem less likely. Note, however, 
that no polarimetric data exist for this MSP to conclusively confirm or rule out 
the presence of a core beam.

From our models we have estimated $f_{\Omega}$ (described in Section
2.4) for each model and provide estimated uncertainties (see Table \ref{modTab}).  
In all cases the predicted $f_{\Omega}$ is
less than 1, although relatively close. This leads to corrected gamma-ray efficiencies of 
$\eta = \left( 0.8 \pm 0.1 \pm 0.1 \right) \times \left( d / 1.8\ \rm kpc\right)^2$ for the TPC model 
and $\eta = \left( 0.9_{-0.3 - 0.3}^{+0.1 + 0.1} \right) \times \left( d / 1.8\ \rm kpc\right)^2$ for the OG model.

\begin{table*}
\caption{Best-fit parameters from light curve modeling (see Section \ref{model}). 
For the gap width parameters the maximum size is 0.10. Values of 0.0 are unphysical 
and should be taken to mean that the best-fit width is less than our resolution of 0.025. 
For the OG models the width of the emitting gap is constrained to be no more than 
one-half the accelerating gap size.\label{modTab}}
\begin{footnotesize}
\begin{tabular}{ccccccccc}
\hline
Gamma-ray model & Radio model & $\alpha$ ($\degr$) & $\zeta$ ($\degr$) & $w_{acc} (\theta_{\rm PC})$ & $w_{em} (\theta_{\rm PC})$ & $-\ln(\mathcal{L})$ & $n_{\rm free}$ & $f_{\Omega}$\\
\hline
TPC & Hollow Cone & $52^{+5}_{-6}$ & $76^{+4}_{-3}$ & $0.10^{+0.025}_{-0.075}$ & -- & 142.2 & 54 & $0.81^{+0.06}_{-0.09}$\\
 & Core Only & $49^{+12}_{-7}$ & $78^{+2}_{-7}$ & $0.10^{+0}_{0.05}$ & -- & 138.3 & 54 & $0.78^{+0.17}_{-0.07}$ \\
 & Core + Cone & $50^{+5}_{-8}$ & $77^{+3}_{-3}$ & $0.10^{+0}_{0.05}$& -- & 136.3 & 53 & $0.78^{+0.09}_{-0.07}$ \\
\hline
OG & Hollow Cone & $56^{+2}_{-14}$ & $79 \pm 1$ & $0.0^{+0.075}_{-0}$ & $0.0^{+0.025}_{-0}$ & 185.3 & 53 & 0.92$^{+0.01}_{-0.26}$\\
 & Core Only & $45^{+9}_{-3}$ & $78^{+2}_{-1}$ & $0.0^{+0.025}_{-0}$ & $0.0^{+0.025}_{-0}$ & 180.6 & 53 & $0.79^{+0.13}_{-0.05}$ \\
 & Core + Cone & $45^{+12}_{-3}$ & $78^{+2}_{-1}$ & $0.0^{+0.025}_{-0}$ & $0.0^{+0.025}_{-0}$ & 179.1 & 52 & $0.79^{+0.15}_{-0.05}$ \\
\hline
\end{tabular}
\end{footnotesize}
\end{table*}

\subsection{Component masses}
\label{sec:masses}

\begin{figure}
\includegraphics[scale=0.45]{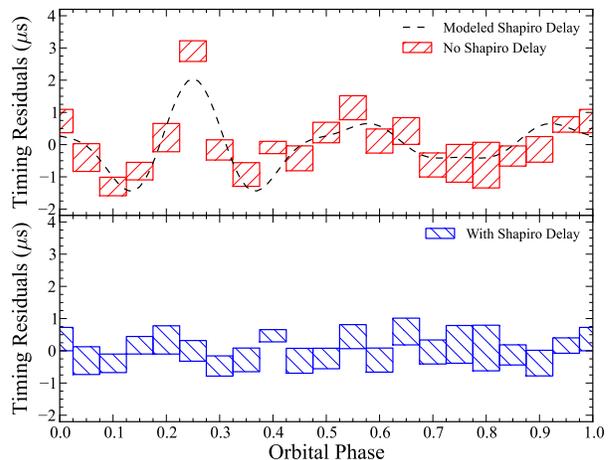}
\caption{Arrival time residuals for the 0.3 GHz Arecibo TOAs of PSR~J2043+1711, 
as a function of orbital phase. The timing residuals were binned in phase, with 20 bins per orbit. 
TOA uncertainties were taken into account when binning the timing residuals. \textit{Bottom:} 
residuals obtained with the full timing model listed in Table \ref{ephem}. 
\textit{Top:} residuals for best-fit orbital model that does not take into account the Shapiro delay. 
The dashed line shows the theoretical prediction for the detectable part of the Shapiro delay
(which is not absorbed by the fitting of the Keplerian parameters $x$
and $e$) given by Eq. (19) in \citet{Freire2010} and the $h_3$ and $\varsigma$
parameters in Table \ref{ephem}. \label{shapiro}}
\end{figure}

\begin{figure*}
\includegraphics[scale=0.6]{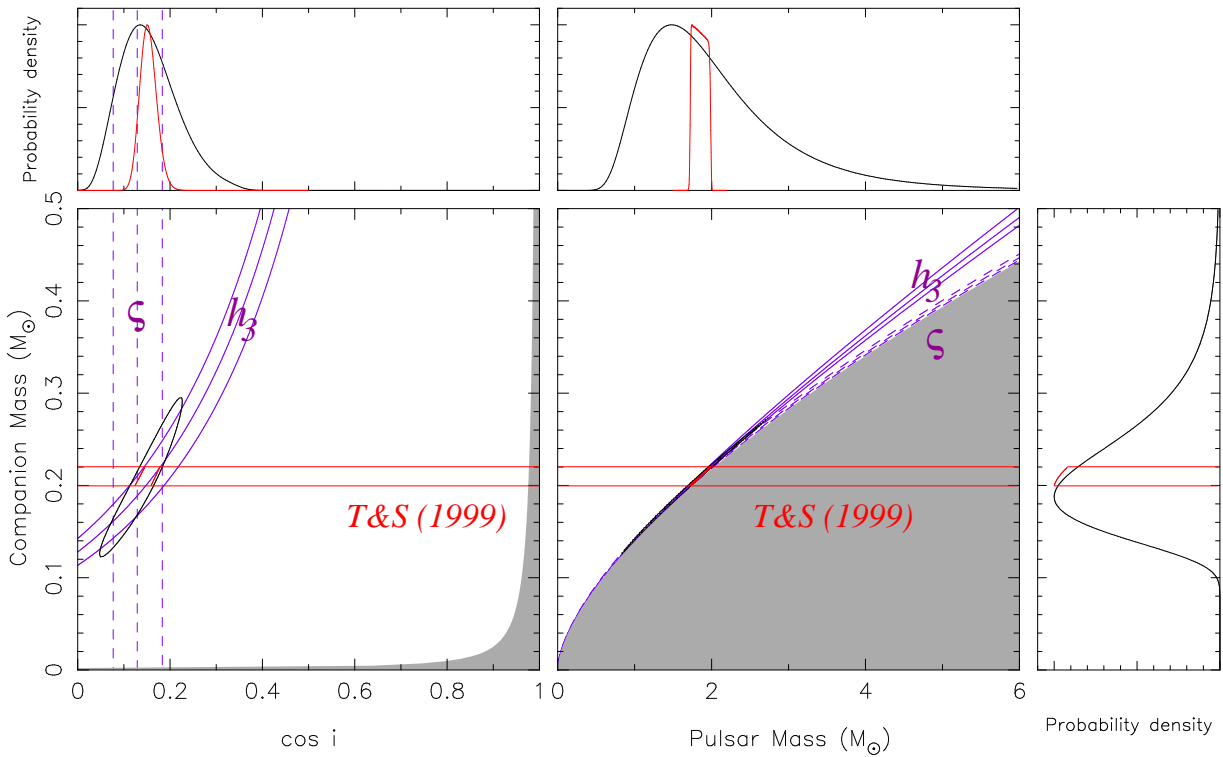}
\caption{Constraints on some physical parameters (companion mass $m_c$,
  pulsar mass $m_p$ and orbital inclination $i$) of the
PSR~J2043+1711 binary system. The purple curves enclose regions
consistent with the nominal values and 1$\sigma$ uncertainties
of the Shapiro delay parameters $h_3$ (solid) and $\varsigma$ (dashed).
{\em Left}: $m_c$-$\cos i$ plot. The gray region is excluded by the
condition $m_p > 0$. The black solid curve is a contour level of the
2D probability density function (PDF) that encloses 68.3\%
of the total probability.
{\em Right}: $m_c$-$m_p$ plot. The gray region is excluded by the
condition $\sin i \leq 1$. The black solid curve encloses 68.3\% of 
the total probability in this region of the plane, and is not a translation of 
the black contour curve in the left-hand plot.
{\em Top, right marginal plots:} the solid black lines show the 1D probability distribution functions 
for $\cos i$, $m_p$ and $m_c$. For all plots, the red contours
represent sections of the same 2D PDF for which
$0.20\,<\,m_c\,<\,0.22\, M_{\sun}$  according to the \citet{Tauris1999} 
model based on the orbital period measured for PSR~J2043+1711. \label{massmass}}
\end{figure*}

From the projected semi-major axis $x$ and the orbital period $P_b$ 
we calculate the mass function $f(m_p, m_c) = (m_c \sin i)^3 / (m_p + m_c)^2 = 
(4 \pi^2 c^3 x^3) / (G M_{\sun} P_b^2) \sim 2.1 \times 10^{-3} M_{\sun}$, 
where $m_p$, $m_c$ and $i$ are the pulsar mass, the companion mass and the orbital inclination. 
Assuming a pulsar mass $m_p$ of 1.4 $M_{\sun}$, we derive 
a minimum companion mass of $m_c\,>\,0.173\ M_{\sun}$. 
The mass of the companion could be significantly higher for much lower orbital
inclinations. However, that possibility can be excluded by our
detection of relativistic time delays in the TOAs caused by spacetime
curvature in the vicinity of the companion star, commonly known as the
Shapiro delay \citep{Shapiro1964}; these are displayed in Figure~\ref{shapiro}. 
This detection has high significance: the orthometric amplitude ($h_3$) 
measurement is nine times larger than the 1$\sigma$ uncertainty. This detection implies 
that the orbital inclination must be high. The constraints on $m_c$, $m_p$ 
and $\sin i$ introduced by our detection of the Shapiro delay are depicted in Figure~\ref{massmass}.
The mass function and range of companion mass values suggest that the companion
star is likely to be a He-type white dwarf (WD).

The second Shapiro delay parameter, the orthometric 
ratio ($\varsigma$), is not yet precise enough to determine 
astrophysically meaningful values for $m_c$, $m_p$ and $\sin i$.
We can, nevertheless, make an estimate of $m_p$ if we assume that
the \citet{Tauris1999} relation between $m_c$ and
$P_b$ applies, as it does for all known MSP/He-type-WD systems 
with precisely measured masses \citep[see e.g. Figure 2 of][]{vanKerkwijk2005}. 
For the orbital period of PSR~J2043+1711, we derive 
$0.20\,<\,m_c\,<\,0.22\,M_{\sun}$, which is very close to the
best value derived from the Shapiro delay 
measurements as can be seen from Figure \ref{massmass}. 
In this case, the existing constraints imply $1.7 < m_p < 2.0$ M$_{\sun}$ and
$i\,=\,81.3 \pm 1.0$ degrees (that is, a nearly edge-on configuration). Thus
the mass of this pulsar appears to be located between the masses of
PSR~J1903+0327 \citep{Freire2011} and PSR~J1614$-$2230 \citep{Demorest2010}, 
which define the high end of the millisecond pulsar mass distribution.

\subsection{Constraints on the distance and the moment of inertia}
\label{sec:mi}

With the current set of TOAs we were able to measure the pulsar's
proper motion, finding $\mu_\alpha \cos(\delta)= -7 \pm 1$ mas yr$^{-1}$ and
$\mu_\delta = -11 \pm 2$ mas yr$^{-1}$, corresponding to a total
proper motion of $\mu_T = 13 \pm 2$ mas yr$^{-1}$. 
For a distance of 1.8 kpc, this gives a transverse velocity 
$V_T \sim 110$~km~s$^{-1}$, a value that is relatively typical 
among Galactic disk millisecond pulsars \citep[see e.g.][]{Hobbs2005}. 

This transverse motion induces a constantly changing Doppler shift
first noted by \citet{Shklovskii1970}, which makes the apparent
$\dot{P}$ value greater than the intrinsic one by $\dot{P}_s \sim 2.43
\times 10^{-21}$~s$^{-1}$~$P d \mu_T^2$, where $P$ is the rotational
period and $d$ is the distance. The NE2001 model of Galactic free
electron density distribution places the pulsar at $1.8 \pm 0.4$ kpc
\citep{NE2001}. For this distance and proper motion, $\dot
P_s$ is found to represent $\sim$30\% of the measured (apparent) $\dot
P$ value. The Shklovskii-corrected period derivative $\dot P_{corr}$
and the derived pulsar properties are listed in Table \ref{ephem}.

\begin{figure}
\includegraphics[scale=0.45]{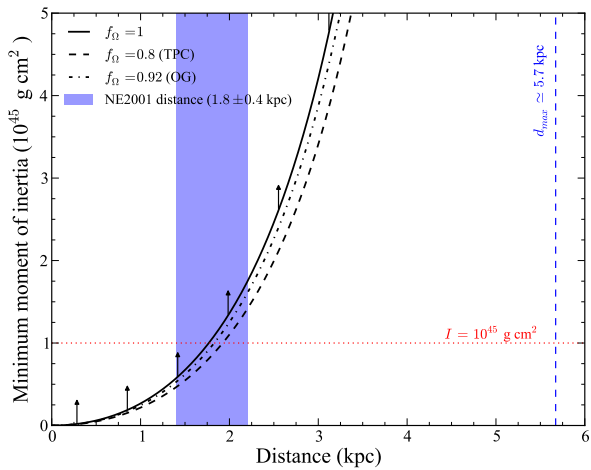}
\caption{Upper limits on the distance as a function of the moment of
  inertia, imposed by the measured gamma-ray energy flux, an 100\%
  upper limit for the gamma-ray efficiency and different values for $f_{\Omega}$, 
  including best-fit values obtained from radio and gamma-ray light curve 
  modeling (see Section \ref{model} and Table \ref{modTab}). Normally $I$
  is assumed to be $10^{45}\, \rm g \,cm^{2}$, this would imply a maximum
  possible distance of about 1.8 kpc for $f_{\Omega} = 1$. If the
  distance is measured precisely, these limiting curves can be
  re-interpreted as lower limits on the moment of inertia. At the
  maximum possible distance
  $d_{\rm max}$ (to be refined as the measurement of the proper motion
  improves) the whole observed $\dot{P}_{\mathrm obs}$ would be due to
  kinematic effects, implying $\dot{P}_{\mathrm corr} = 0$. Therefore,
  for the energy loss of the pulsar $\dot{E} = 4 \pi^2 I
  \dot{P}_{\mathrm corr}/P^3$ to account for the gamma-ray luminosity,
  $I$ would have to be infinite. \label{inertia}}
\end{figure}

An upper limit on the distance can be derived 
by assuming that the Shklovskii effect accounts for all of the
apparent $\dot P$ and that the currently measured proper motion is
correct. Doing so, we find $d_{max} = 5.7$ kpc (see
Figure~\ref{inertia}). It is clear, however, that the pulsar must
be at a distance from
the Solar System that is significantly smaller than $d_{max}$.
The reason for this has already been briefly mentioned in
Section~\ref{gammaanalysis}, namely the very high implied gamma-ray efficiency: 
for distances larger than 1.8 kpc, an efficiency larger than 100\% 
is required to produce the gamma-ray flux detected by the LAT. 

We can derive more realistic upper limits for $d$ based on our
\emph{Fermi} LAT measurement of the energy density $G$ and imposing
the condition $\eta < 1$. A comparison of the expressions for
$\dot{E}$ and $L_{\gamma}$ then yields the following inequality:
\begin{equation}
I > \frac{G f_{\Omega} d^2 P^3}{\pi \dot{P}_{\rm corr}(d, \mu_T)}.
\label{eq:distance_I}
\end{equation}
For the current assumptions of $I = 10^{45}$ g cm$^{2}$ and
$f_{\Omega} = 1$, this inequality only holds for distances
smaller than 1.8 kpc (Figure~\ref{inertia}). For the best-fit 
$f_{\Omega}$ factors obtained from the modeling of radio and 
gamma-ray light curves of PSR~J2043+1711 with the TPC 
and OG models and with the hollow-cone beam radio model, 
the inequality holds for $d < 1.92$ and 1.83 kpc, respectively.
The larger distances within these ranges are
comparable to the distance predicted by the NE2001 model.

Improved radio timing of this pulsar might eventually allow a
precise measurement of the timing parallax, and therefore 
of the distance $d$, which according to Eq.~\ref{eq:distance_I}
provides a direct lower limit for $I$.
Again, such limits are displayed in Figure~\ref{inertia}.
As was mentioned in Section~\ref{gammaanalysis}, the NE2001 model
likely overestimates the distance. However, 
it is clear that if the model is correct or the pulsar is
farther away, then its moment of inertia must be large. 
A high lower limit on the moment of inertia would help constrain 
equations of state (EoSs) for super-dense matter \citep{Worley2008}, especially 
if combined with a precise measurement for the mass of the
pulsar, another likely consequence of the improved timing.

\section{Conclusion and prospects}
\label{conclusion}

We have reported the discovery of an MSP with the Nan\c cay Radio 
Telescope at the position of an unassociated \emph{Fermi} source,
PSR~J2043+1711. The pulsar is the third MSP to be discovered at Nan\c cay 
in a \emph{Fermi} source, after PSRs J2017+0603 and 
J2302+4442 \citep{Cognard2011}. 
The radio pulsar is found to be responsible for the gamma-ray 
emission observed by \emph{Fermi}, and its properties (rotational period, 
spin-down luminosity, distance, gamma-ray light curve and gamma-ray spectrum) 
are relatively common among known gamma-ray MSPs. 

Of the pulsars discovered in \emph{Fermi} unassociated sources
that have been published to date 
\citep{Cognard2011,Ransom2011,Keith2011}, PSRs J2043+1711 and
J2017+0603 are the two systems with the best timing precision.
This happens because they have very sharp features in their pulse profiles.
Both objects are relatively faint, but they are well within the region
of the sky detectable by the Arecibo 305-m telescope, which greatly
compensates for the small flux density. Furthermore, both pulsars appear
to be members of very stable MSP-WD binaries. 
For this reason, they have been included in the
pulsar timing array (PTA), which is now being used in
a collective effort to detect low-frequency gravitational waves
\citep[e.g.][]{Hobbs2010}.

Given this high timing precision, continued
Arecibo timing might provide a precise distance measurement
which, given the observed gamma-ray energy flux, would
provide a lower limit for the moment of inertia of this MSP.
This distance is likely to be significantly smaller than the
estimate provided by the NE2001 electron model of the Galaxy, in
which case the parallax should be easier to measure but the
lower limit on $I$ would not be constraining.
If the distance is comparable to the DM prediction or
larger, then the parallax will be more
difficult to measure, but a low upper limit for the parallax would
allow us to derive a high lower limit for the moment of inertia.
This might constrain the EoS of
neutron matter at densities above that of the atomic nucleus -- 
some predict $I < 1.7 \times 10^{45}$ g cm$^{2}$ \citep{Worley2008}; 
measuring a larger lower limit for $I$ would exclude such EoSs.

Continued timing of PSR~J2043+1711 will also improve the
measurement of the Shapiro delay, providing precise estimates
of the masses of the components of the system. Combining this precise
mass with a lower limit on the moment of inertia could
provide a stringent constraint of the EoS.

\section*{Acknowledgements}

We would like to thank the anonymous referee for the valuable 
suggestions to this paper. 

The \textit{Fermi} LAT Collaboration acknowledges generous ongoing
support from a number of agencies and institutes that have supported
both the development and the operation of the LAT as well as
scientific data analysis. These include the National Aeronautics and
Space Administration and the Department of Energy in the United
States, the Commissariat \`a l'Energie Atomique and the Centre
National de la Recherche Scientifique / Institut National de Physique
Nucl\'eaire et de Physique des Particules in France, the Agenzia
Spaziale Italiana and the Istituto Nazionale di Fisica Nucleare in
Italy, the Ministry of Education, Culture, Sports, Science and
Technology (MEXT), High Energy Accelerator Research Organization (KEK)
and Japan Aerospace Exploration Agency (JAXA) in Japan, and the
K.~A.~Wallenberg Foundation, the Swedish Research Council and the
Swedish National Space Board in Sweden.

Additional support for science analysis during the operations phase is
gratefully acknowledged from the Istituto Nazionale di Astrofisica in
Italy and the Centre National d'\'Etudes Spatiales in France.

The Nan\c cay Radio Observatory is operated by the Paris Observatory,
associated with the French Centre National de la Recherche
Scientifique (CNRS). The Arecibo Observatory is part of the National
Astronomy and Ionosphere Center (NAIC), a national research center
operated by Cornell University under a cooperative agreement with the
National Science Foundation. The Green Bank Telescope is operated by
the National Radio Astronomy Observatory, a facility of the National
Science Foundation operated under cooperative agreement by Associated
Universities, Inc. The Westerbork Synthesis Radio Telescope is operated by 
Netherlands Foundation for Radio Astronomy, ASTRON.

\appendix

\section{Radio search observations with the Nan\c cay Radio Telescope}

To avoid redundant observations with other instruments, and to inform 
the community of existing data samples, Table \ref{obslog} lists  all pulsar searches 
of \emph{Fermi} error ellipses made with the NRT to date. 

For each observation, Table \ref{obslog} lists the observation time $T_{obs}$, the 
sky temperature in the corresponding direction $T_{sky}$, and the minimum 
detectable flux density $S_{min}$. The sky temperature $T_{sky}$ corresponds to the 
contribution from the Galactic synchrotron component, calculated 
by scaling the 408 MHz map of \citet{Haslam1982} to the observing frequency 
of 1.4 GHz with a spectral index of $-2.6$. The quantity $S_{min}$ was 
estimated using the modified radiometer equation \citep[see e.g.][]{Handbook}, 
with $G = 1.4$ K Jy$^{-1}$, $n_p = 2$, $\beta = 1.05$, 
$T_{sys} = T_{rec} + T_{sky}$ with $T_{rec} = 35$ K 
(note that this receiver temperature includes the 2.7 K temperature from 
the cosmic microwave background), $\Delta f = 128$ MHz, 
and assuming $( S / N )_{\rm min} = 5$, and $W = 0.1 \times P$. 
Also listed in the Table are telescope pointing directions, 
and the offsets from the corresponding 2FGL sources. 

For each source, Table \ref{obslog} gives the semi-major axis of 
the 95\% confidence ellipse ($\theta_{95}$), the curvature significance (``Signif\_Curve''), 
the variability index (``Variability\_Index''), and the name of the 
identified or likely associated source (``Assoc.''), if any. Details on the 
determination of these parameters and on the source association procedure can 
be found in \citet{Fermi2FGL}.

Of the sources observed with the NRT and for which no radio pulsations have 
been observed to date, the following have curvature significances above 4$\sigma$ and 
variability indices smaller than 41.6, making them good pulsar candidates: 
2FGL~J0224.0+6204, J0734.6$-$1558, 
J1120.0$-$2204, J1311.7$-$3429, J1625.2$-$0020, and J2339.6$-$0532. 
With the exception of 2FGL~J0734.6$-$1558, in which a gamma-ray pulsar has been 
discovered through blind searches of the \emph{Fermi} LAT data \citep{SazParkinson2011}, 
these sources remain unassociated. Multi-wavelength studies 
might help determine their natures. Optical and X-ray observations of 
2FGL~J2339.6$-$0532 showed that it is likely to be powered by a millisecond pulsar 
in a black-widow system \citep{Romani2011,Kong2012}.

\bibliographystyle{mnras}

\bibliography{J2043+1711}

\bsp

\onecolumn

\begin{center}
\begin{scriptsize}
\begin{landscape}

\begin{longtable}{@{\extracolsep{-6pt}}ccccccccccccccc}
\caption{\small \emph{Fermi} LAT sources searched for radio pulsars at the Nan\c cay Radio Telescope. 
Definitions of parameters and details on the calculation of the sky temperature $T_{sky}$ 
and the minimum detectable flux density $S_{min}$ for each observation are given in the text. Note that 
Right ascensions and declinations refer to the telescope pointing directions, and not 
necessarily to the locations of the \emph{Fermi} LAT sources.\label{obslog}}\\

\hline
2FGL Name & 1FGL Name & $l$ & $b$ & $\theta_{95}$ & Date & R.A. & Decl. & Offset & $T_{obs}$ & $T_{sky}$ & 
$S_{min}$ & Signif\_Curve & Variability\_Index & Assoc. \\
 & & (deg) & (deg) & (arcmin) & & (J2000.0) & (J2000.0) & (arcmin) & (min) & (K) & ($\mu$Jy) & & & \\
\hline
\endfirsthead

Continued from previous page.\\
\hline
2FGL Name & 1FGL Name & $l$ & $b$ & $\theta_{95}$ & Date & R.A. & Decl. & Offset & $T_{obs}$ & $T_{sky}$ & 
$S_{min}$ & Signif\_Curve & Variability\_Index & Assoc. \\
 & & (deg) & (deg) & (arcmin) & & (J2000.0) & (J2000.0) & (arcmin) & (min) & (K) & ($\mu$Jy) & & & \\
\hline
\endhead

\hline
\endfoot

\hline
\endlastfoot

J0031.0+0724 & J0030.7+0724 & 114.10 & $-$55.11 & 7.2 & 2010-01-08 & $00^{\rmn{h}} 30^{\rmn{m}} 45^{\rmn{s}}$ & $+07\degr 22\arcmin 34\arcsec$ & 5.7 & 58.5 & 0.9 & 47 & 0.8 & 20.9 &  \\
J0102.7+5827 & J0102.8+5827 & 124.42 & $-$4.38 & 3.6 & 2009-12-17 & $01^{\rmn{h}} 02^{\rmn{m}} 50^{\rmn{s}}$ & $+58\degr 27\arcmin 18\arcsec$ & 0.8 & 66.0 & 2.0 & 46 & 3.1 & 119.3 & TXS 0059+581 \\
 &  &  &  &  & 2009-12-23 & $01^{\rmn{h}} 02^{\rmn{m}} 50^{\rmn{s}}$ & $+58\degr 27\arcmin 18\arcsec$ & 0.8 & 66.0 & 2.0 & 46 &  &  &  \\
 &  &  &  &  & 2010-01-09 & $01^{\rmn{h}} 02^{\rmn{m}} 50^{\rmn{s}}$ & $+58\degr 27\arcmin 18\arcsec$ & 0.8 & 66.0 & 2.0 & 46 &  &  &  \\
 &  &  &  &  & 2010-01-10 & $01^{\rmn{h}} 02^{\rmn{m}} 50^{\rmn{s}}$ & $+58\degr 27\arcmin 18\arcsec$ & 0.8 & 78.0 & 2.0 & 42 &  &  &  \\
J0118.8$-$2142 & J0118.7$-$2137 & 173.46 & $-$81.73 & 2.9 & 2010-04-18 & $01^{\rmn{h}} 18^{\rmn{m}} 43^{\rmn{s}}$ & $-21\degr 37\arcmin 39\arcsec$ & 5.4 & 57.2 & 0.9 & 48 & 4.1 & 257.2 & PKS 0116$-$219 \\
 &  &  &  &  & 2010-05-14 & $01^{\rmn{h}} 18^{\rmn{m}} 43^{\rmn{s}}$ & $-21\degr 37\arcmin 39\arcsec$ & 5.4 & 54.6 & 0.9 & 49 &  &  &  \\
J0124.5$-$0621 &  J0124.6$-$0616 & 145.21 & $-$67.79 & 12.1 & 2011-03-30 & $01^{\rmn{h}} 24^{\rmn{m}} 35^{\rmn{s}}$ & $-06\degr 21\arcmin 51\arcsec$ & 0.2 & 29.7 & 0.9 & 66 & 2.7 & 28.8 & PMN J0124$-$0624 \\
 &  &  &  &  & 2011-04-01 & $01^{\rmn{h}} 24^{\rmn{m}} 35^{\rmn{s}}$ & $-06\degr 21\arcmin 51\arcsec$ & 0.2 & 39.7 & 0.9 & 58 &  &  &  \\
 &  &  &  &  & 2011-04-30 & $01^{\rmn{h}} 24^{\rmn{m}} 35^{\rmn{s}}$ & $-06\degr 21\arcmin 51\arcsec$ & 0.2 & 49.7 & 0.9 & 51 &  &  &  \\
J0127.2+0324 &  J0127.0+0322 & 140.12 & $-$58.26 & 5.8 & 2011-03-26 & $01^{\rmn{h}} 27^{\rmn{m}} 17^{\rmn{s}}$ & $+03\degr 25\arcmin 11\arcsec$ & 0.8 & 51.0 & 0.9 & 51 & 0.1 & 31.9 & NVSS J012713+032259 \\
 &  &  &  &  & 2011-03-29 & $01^{\rmn{h}} 27^{\rmn{m}} 17^{\rmn{s}}$ & $+03\degr 25\arcmin 11\arcsec$ & 0.8 & 51.0 & 0.9 & 51 &  &  &  \\
 &  &  &  &  & 2011-05-07 & $01^{\rmn{h}} 27^{\rmn{m}} 17^{\rmn{s}}$ & $+03\degr 25\arcmin 11\arcsec$ & 0.8 & 52.0 & 0.9 & 50 &  &  &  \\
J0131.1+6121 & J0131.2+6121 & 127.67 & $-$1.15 & 2.5 & 2009-12-22 & $01^{\rmn{h}} 31^{\rmn{m}} 17^{\rmn{s}}$ & $+61\degr 21\arcmin 42\arcsec$ & 1.1 & 66.0 & 2.4 & 46 & 2.4 & 58.1 & 1RXS J013106.4+612035 \\
 &  &  &  &  & 2009-12-30 & $01^{\rmn{h}} 31^{\rmn{m}} 17^{\rmn{s}}$ & $+61\degr 21\arcmin 42\arcsec$ & 1.1 & 58.6 & 2.4 & 49 &  &  &  \\
 &  &  &  &  & 2009-12-31 & $01^{\rmn{h}} 31^{\rmn{m}} 17^{\rmn{s}}$ & $+61\degr 21\arcmin 42\arcsec$ & 1.1 & 34.7 & 2.4 & 64 &  &  &  \\
 &  &  &  &  & 2010-09-25 & $01^{\rmn{h}} 31^{\rmn{m}} 17^{\rmn{s}}$ & $+61\degr 21\arcmin 42\arcsec$ & 1.1 & 53.0 & 2.4 & 52 &  &  &  \\
 &  &  &  &  & 2010-12-02 & $01^{\rmn{h}} 31^{\rmn{m}} 08^{\rmn{s}}$ & $+61\degr 20\arcmin 45\arcsec$ & 0.6 & 31.0 & 2.4 & 68 &  &  &  \\
J0137.7+5811 & J0137.8+5814 & 129.02 & $-$4.12 & 6.0 & 2010-01-12 & $01^{\rmn{h}} 37^{\rmn{m}} 46^{\rmn{s}}$ & $+58\degr 14\arcmin 07\arcsec$ & 2.2 & 34.7 & 2.0 & 63 & 0.7 & 15.3 & 1RXS J013748.0+581422 \\
 &  &  &  &  & 2010-01-13 & $01^{\rmn{h}} 37^{\rmn{m}} 46^{\rmn{s}}$ & $+58\degr 14\arcmin 07\arcsec$ & 2.2 & 66.0 & 2.0 & 46 &  &  &  \\
 &  &  &  &  & 2010-01-20 & $01^{\rmn{h}} 37^{\rmn{m}} 46^{\rmn{s}}$ & $+58\degr 14\arcmin 07\arcsec$ & 2.2 & 32.5 & 2.0 & 65 &  &  &  \\
 &  &  &  &  & 2010-12-01 & $01^{\rmn{h}} 38^{\rmn{m}} 04^{\rmn{s}}$ & $+58\degr 13\arcmin 12\arcsec$ & 2.7 & 23.5 & 2.0 & 77 &  &  &  \\
 &  &  &  &  & 2010-12-14 & $01^{\rmn{h}} 38^{\rmn{m}} 04^{\rmn{s}}$ & $+58\degr 13\arcmin 12\arcsec$ & 2.7 & 39.7 & 2.0 & 59 &  &  &  \\
J0224.0+6204 &  J0224.0+6201c & 133.55 & 1.13 & 3.4 & 2011-04-08 & $02^{\rmn{h}} 24^{\rmn{m}} 28^{\rmn{s}}$ & $+62\degr 02\arcmin 59\arcsec$ & 3.1 & 47.4 & 3.0 & 56 & 7.2 & 20.1 &  \\
J0250.7+5631 & J0251.5+5634 & 138.87 & $-$2.61 & 7.1 & 2009-12-22 & $02^{\rmn{h}} 50^{\rmn{m}} 47^{\rmn{s}}$ & $+56\degr 31\arcmin 50\arcsec$ & 0.3 & 60.0 & 2.1 & 48 & 0.6 & 28.0 & NVSS J025047+562935 \\
 &  &  &  &  & 2009-12-30 & $02^{\rmn{h}} 50^{\rmn{m}} 47^{\rmn{s}}$ & $+56\degr 31\arcmin 50\arcsec$ & 0.3 & 55.1 & 2.1 & 50 &  &  &  \\
 &  &  &  &  & 2009-12-31 & $02^{\rmn{h}} 50^{\rmn{m}} 47^{\rmn{s}}$ & $+56\degr 31\arcmin 50\arcsec$ & 0.3 & 32.5 & 2.1 & 66 &  &  &  \\
J0334.3$-$3728 & J0334.4$-$3727 & 240.22 & $-$54.36 & 3.1 & 2009-11-16 & $03^{\rmn{h}} 34^{\rmn{m}} 32^{\rmn{s}}$ & $-37\degr 27\arcmin 43\arcsec$ & 2.3 & 57.6 & 0.6 & 47 & 2.7 & 90.3 & PMN J0334$-$3725 \\
 &  &  &  &  & 2010-01-14 & $03^{\rmn{h}} 34^{\rmn{m}} 32^{\rmn{s}}$ & $-37\degr 27\arcmin 43\arcsec$ & 2.3 & 60.0 & 0.6 & 46 &  &  &  \\
J0434.1$-$2014 & J0434.1$-$2018 & 217.85 & $-$38.94 & 10.2 & 2009-12-15 & $04^{\rmn{h}} 34^{\rmn{m}} 13^{\rmn{s}}$ & $-20\degr 18\arcmin 10\arcsec$ & 4.0 & 52.2 & 0.7 & 50 & 1.1 & 42.4 & TXS 0431$-$203 \\
 &  &  &  &  & 2010-04-14 & $04^{\rmn{h}} 34^{\rmn{m}} 13^{\rmn{s}}$ & $-20\degr 18\arcmin 10\arcsec$ & 4.0 & 19.7 & 0.7 & 81 &  &  &  \\
 &  &  &  &  & 2010-04-23 & $04^{\rmn{h}} 34^{\rmn{m}} 13^{\rmn{s}}$ & $-20\degr 18\arcmin 10\arcsec$ & 4.0 & 13.5 & 0.7 & 98 &  &  &  \\
 &  &  &  &  & 2010-04-24 & $04^{\rmn{h}} 34^{\rmn{m}} 13^{\rmn{s}}$ & $-20\degr 18\arcmin 10\arcsec$ & 4.0 & 13.5 & 0.7 & 98 &  &  &  \\
 &  &  &  &  & 2011-04-05 & $04^{\rmn{h}} 34^{\rmn{m}} 13^{\rmn{s}}$ & $-20\degr 18\arcmin 10\arcsec$ & 4.0 & 44.7 & 0.7 & 54 &  &  &  \\
J0523.3$-$2530 & J0523.5$-$2529 & 228.23 & $-$29.84 & 4.0 & 2009-11-08 & $05^{\rmn{h}} 23^{\rmn{m}} 32^{\rmn{s}}$ & $-25\degr 30\arcmin 26\arcsec$ & 3.0 & 32.2 & 0.6 & 63 & 2.1 & 22.7 &  \\
 &  &  &  &  & 2009-11-17 & $05^{\rmn{h}} 23^{\rmn{m}} 32^{\rmn{s}}$ & $-25\degr 30\arcmin 26\arcsec$ & 3.0 & 24.7 & 0.6 & 72 &  &  &  \\
J0714.0+1933 & J0714.0+1935 & 197.68 & 13.61 & 2.6 & 2009-11-06 & $07^{\rmn{h}} 14^{\rmn{m}} 02^{\rmn{s}}$ & $+19\degr 34\arcmin 39\arcsec$ & 0.8 & 38.5 & 1.2 & 59 & 5.1 & 313.4 & MG2 J071354+1934 \\
J0734.6$-$1558 &  J0734.7$-$1557 & 232.04 & 2.01 & 3.4 & 2010-09-07 & $07^{\rmn{h}} 34^{\rmn{m}} 45^{\rmn{s}}$ & $-15\degr 59\arcmin 18\arcsec$ & 1.4 & 66.0 & 1.4 & 45 & 5.9 & 24.5 & LAT PSR J0734$-$1559 \\
J0928.8$-$3530 & J0929.0$-$3531 & 263.02 & 11.23 & 7.7 & 2010-02-08 & $09^{\rmn{h}} 28^{\rmn{m}} 56^{\rmn{s}}$ & $-35\degr 32\arcmin 16\arcsec$ & 2.1 & 23.5 & 1.4 & 76 & 2.8 & 23.2 &  \\
J1120.0$-$2204 & J1119.9$-$2205 & 276.50 & 36.05 & 3.8 & 2010-01-14 & $11^{\rmn{h}} 19^{\rmn{m}} 54^{\rmn{s}}$ & $-22\degr 05\arcmin 06\arcsec$ & 1.5 & 44.7 & 0.6 & 54 & 5.5 & 25.9 &  \\
 &  &  &  &  & 2010-01-17 & $11^{\rmn{h}} 19^{\rmn{m}} 54^{\rmn{s}}$ & $-22\degr 05\arcmin 06\arcsec$ & 1.5 & 36.0 & 0.6 & 60 &  &  &  \\
 &  &  &  &  & 2010-07-18 & $11^{\rmn{h}} 19^{\rmn{m}} 54^{\rmn{s}}$ & $-22\degr 05\arcmin 06\arcsec$ & 1.5 & 60.0 & 0.6 & 46 &  &  &  \\
 &  &  &  &  & 2011-03-29 & $11^{\rmn{h}} 19^{\rmn{m}} 58^{\rmn{s}}$ & $-22\degr 05\arcmin 26\arcsec$ & 0.8 & 56.0 & 0.6 & 48 &  &  &  \\
 &  &  &  &  & 2011-03-30 & $11^{\rmn{h}} 19^{\rmn{m}} 58^{\rmn{s}}$ & $-22\degr 05\arcmin 26\arcsec$ & 0.8 & 57.2 & 0.6 & 47 &  &  &  \\
 &  &  &  &  & 2011-04-22 & $11^{\rmn{h}} 19^{\rmn{m}} 58^{\rmn{s}}$ & $-22\degr 05\arcmin 26\arcsec$ & 0.8 & 47.2 & 0.6 & 52 &  &  &  \\
J1120.4+0710 & J1120.4+0710 & 251.53 & 60.61 & 5.3 & 2009-11-27 & $11^{\rmn{h}} 20^{\rmn{m}} 30^{\rmn{s}}$ & $+07\degr 10\arcmin 22\arcsec$ & 0.5 & 27.2 & 0.7 & 69 & 2.8 & 26.6 & MG1 J112039+0704 \\
 &  &  &  &  & 2009-12-03 & $11^{\rmn{h}} 20^{\rmn{m}} 30^{\rmn{s}}$ & $+07\degr 10\arcmin 22\arcsec$ & 0.5 & 66.0 & 0.7 & 44 &  &  &  \\
 &  &  &  &  & 2010-07-13 & $11^{\rmn{h}} 20^{\rmn{m}} 30^{\rmn{s}}$ & $+07\degr 10\arcmin 22\arcsec$ & 0.5 & 66.0 & 0.7 & 44 &  &  &  \\
 &  &  &  &  & 2010-07-14 & $11^{\rmn{h}} 20^{\rmn{m}} 30^{\rmn{s}}$ & $+07\degr 10\arcmin 22\arcsec$ & 0.5 & 66.0 & 0.7 & 44 &  &  &  \\
J1129.5+3758 & J1129.3+3757 & 175.54 & 69.69 & 8.7 & 2009-11-19 & $11^{\rmn{h}} 29^{\rmn{m}} 23^{\rmn{s}}$ & $+37\degr 58\arcmin 19\arcsec$ & 1.6 & 55.4 & 0.8 & 49 & 0.4 & 28.3 &  \\
 &  &  &  &  & 2010-01-11 & $11^{\rmn{h}} 29^{\rmn{m}} 23^{\rmn{s}}$ & $+37\degr 58\arcmin 19\arcsec$ & 1.6 & 32.9 & 0.8 & 63 &  &  &  \\
 &  &  &  &  & 2010-05-10 & $11^{\rmn{h}} 29^{\rmn{m}} 23^{\rmn{s}}$ & $+37\degr 58\arcmin 19\arcsec$ & 1.6 & 29.7 & 0.8 & 66 &  &  &  \\
 &  &  &  &  & 2010-05-12 & $11^{\rmn{h}} 29^{\rmn{m}} 23^{\rmn{s}}$ & $+37\degr 58\arcmin 19\arcsec$ & 1.6 & 26.1 & 0.8 & 71 &  &  &  \\
 &  &  &  &  & 2010-07-21 & $11^{\rmn{h}} 29^{\rmn{m}} 23^{\rmn{s}}$ & $+37\degr 58\arcmin 19\arcsec$ & 1.6 & 39.7 & 0.8 & 57 &  &  &  \\
 &  &  &  &  & 2010-12-01 & $11^{\rmn{h}} 29^{\rmn{m}} 56^{\rmn{s}}$ & $+37\degr 59\arcmin 05\arcsec$ & 5.0 & 37.2 & 0.8 & 59 &  &  &  \\
 &  &  &  &  & 2010-12-12 & $11^{\rmn{h}} 29^{\rmn{m}} 56^{\rmn{s}}$ & $+37\degr 59\arcmin 05\arcsec$ & 5.0 & 66.0 & 0.8 & 44 &  &  &  \\
J1311.7$-$3429 & J1311.7$-$3429 & 307.69 & 28.20 & 2.0 & 2010-03-30 & $13^{\rmn{h}} 11^{\rmn{m}} 47^{\rmn{s}}$ & $-34\degr 29\arcmin 39\arcsec$ & 0.4 & 60.0 & 1.5 & 48 & 6.3 & 19.1 &  \\
 &  &  &  &  & 2010-05-17 & $13^{\rmn{h}} 11^{\rmn{m}} 47^{\rmn{s}}$ & $-34\degr 29\arcmin 39\arcsec$ & 0.4 & 37.2 & 1.5 & 60 &  &  &  \\
 &  &  &  &  & 2011-05-27 & $13^{\rmn{h}} 11^{\rmn{m}} 47^{\rmn{s}}$ & $-34\degr 29\arcmin 39\arcsec$ & 0.4 & 37.6 & 1.5 & 60 &  &  &  \\
J1340.5$-$0412 & J1340.5$-$0413 & 325.51 & 56.50 & 8.7 & 2010-03-28 & $13^{\rmn{h}} 40^{\rmn{m}} 33^{\rmn{s}}$ & $-04\degr 13\arcmin 56\arcsec$ & 1.5 & 60.0 & 0.9 & 47 & 0.3 & 25.0 &  \\
 &  &  &  &  & 2010-04-13 & $13^{\rmn{h}} 40^{\rmn{m}} 33^{\rmn{s}}$ & $-04\degr 13\arcmin 56\arcsec$ & 1.5 & 60.0 & 0.9 & 47 &  &  &  \\
 &  &  &  &  & 2010-07-21 & $13^{\rmn{h}} 40^{\rmn{m}} 33^{\rmn{s}}$ & $-04\degr 13\arcmin 56\arcsec$ & 1.5 & 32.2 & 0.9 & 64 &  &  &  \\
 &  &  &  &  & 2010-09-07 & $13^{\rmn{h}} 40^{\rmn{m}} 33^{\rmn{s}}$ & $-04\degr 13\arcmin 56\arcsec$ & 1.5 & 41.0 & 0.9 & 57 &  &  &  \\
 &  &  &  &  & 2010-09-09 & $13^{\rmn{h}} 40^{\rmn{m}} 33^{\rmn{s}}$ & $-04\degr 13\arcmin 56\arcsec$ & 1.5 & 41.0 & 0.9 & 57 &  &  &  \\
 &  &  &  &  & 2010-12-02 & $13^{\rmn{h}} 40^{\rmn{m}} 33^{\rmn{s}}$ & $-04\degr 13\arcmin 56\arcsec$ & 1.5 & 53.5 & 0.9 & 50 &  &  &  \\
 &  &  &  &  & 2010-12-04 & $13^{\rmn{h}} 40^{\rmn{m}} 33^{\rmn{s}}$ & $-04\degr 13\arcmin 56\arcsec$ & 1.5 & 31.0 & 0.9 & 65 &  &  &  \\
 &  &  &  &  & 2010-12-08 & $13^{\rmn{h}} 40^{\rmn{m}} 33^{\rmn{s}}$ & $-04\degr 13\arcmin 56\arcsec$ & 1.5 & 27.2 & 0.9 & 69 &  &  &  \\
 &  &  &  &  & 2011-02-11 & $13^{\rmn{h}} 40^{\rmn{m}} 33^{\rmn{s}}$ & $-04\degr 13\arcmin 56\arcsec$ & 1.5 & 52.2 & 0.9 & 50 &  &  &  \\
 &  &  &  &  & 2011-03-22 & $13^{\rmn{h}} 40^{\rmn{m}} 33^{\rmn{s}}$ & $-04\degr 13\arcmin 56\arcsec$ & 1.5 & 51.0 & 0.9 & 51 &  &  &  \\
 &  &  &  &  & 2011-05-09 & $13^{\rmn{h}} 40^{\rmn{m}} 33^{\rmn{s}}$ & $-04\degr 13\arcmin 56\arcsec$ & 1.5 & 44.7 & 0.9 & 54 &  &  &  \\
 &  &  &  &  & 2011-05-13 & $13^{\rmn{h}} 40^{\rmn{m}} 33^{\rmn{s}}$ & $-04\degr 13\arcmin 56\arcsec$ & 1.5 & 46.0 & 0.9 & 53 &  &  &  \\
J1407.4$-$2948 & J1407.9$-$2928 & 322.01 & 30.22 & 21.1 & 2011-03-27 & $14^{\rmn{h}} 07^{\rmn{m}} 25^{\rmn{s}}$ & $-29\degr 48\arcmin 59\arcsec$ & 0.1 & 58.6 & 1.5 & 48 & 3.2 & 17.6 &  \\
 &  &  &  &  & 2011-03-31 & $14^{\rmn{h}} 07^{\rmn{m}} 25^{\rmn{s}}$ & $-29\degr 48\arcmin 59\arcsec$ & 0.1 & 55.0 & 1.5 & 50 &  &  &  \\
 &  &  &  &  & 2011-04-02 & $14^{\rmn{h}} 07^{\rmn{m}} 25^{\rmn{s}}$ & $-29\degr 48\arcmin 59\arcsec$ & 0.1 & 22.2 & 1.5 & 78 &  &  &  \\
 &  &  &  &  & 2011-04-27 & $14^{\rmn{h}} 07^{\rmn{m}} 25^{\rmn{s}}$ & $-29\degr 48\arcmin 59\arcsec$ & 0.1 & 44.7 & 1.5 & 55 &  &  &  \\
 &  &  &  &  & 2011-05-07 & $14^{\rmn{h}} 07^{\rmn{m}} 25^{\rmn{s}}$ & $-29\degr 48\arcmin 59\arcsec$ & 0.1 & 43.5 & 1.5 & 56 &  &  &  \\
 &  &  &  &  & 2011-05-11 & $14^{\rmn{h}} 07^{\rmn{m}} 25^{\rmn{s}}$ & $-29\degr 48\arcmin 59\arcsec$ & 0.1 & 66.0 & 1.5 & 45 &  &  &  \\
J1625.2$-$0020 & J1625.3$-$0019 & 13.92 & 31.83 & 3.4 & 2009-11-11 & $16^{\rmn{h}} 25^{\rmn{m}} 18^{\rmn{s}}$ & $-00\degr 19\arcmin 15\arcsec$ & 1.5 & 54.7 & 1.4 & 50 & 8.3 & 24.6 &  \\
 &  &  &  &  & 2009-11-19 & $16^{\rmn{h}} 25^{\rmn{m}} 18^{\rmn{s}}$ & $-00\degr 19\arcmin 15\arcsec$ & 1.5 & 46.0 & 1.4 & 54 &  &  &  \\
 &  &  &  &  & 2010-01-12 & $16^{\rmn{h}} 25^{\rmn{m}} 18^{\rmn{s}}$ & $-00\degr 19\arcmin 15\arcsec$ & 1.5 & 60.0 & 1.4 & 47 &  &  &  \\
J1642.9+3949 & J1642.5+3947 & 63.48 & 40.95 & 3.2 & 2009-11-04 & $16^{\rmn{h}} 42^{\rmn{m}} 32^{\rmn{s}}$ & $+39\degr 46\arcmin 15\arcsec$ & 6.3 & 66.0 & 0.9 & 45 & 1.3 & 111.7 & 3C 345 \\
 &  &  &  &  & 2009-11-20 & $16^{\rmn{h}} 42^{\rmn{m}} 32^{\rmn{s}}$ & $+39\degr 46\arcmin 15\arcsec$ & 6.3 & 66.0 & 0.9 & 45 &  &  &  \\
J2001.1+4352 & J2001.1+4351 & 79.06 & 7.12 & 1.2 & 2009-11-06 & $20^{\rmn{h}} 01^{\rmn{m}} 11^{\rmn{s}}$ & $+43\degr 52\arcmin 07\arcsec$ & 0.7 & 24.3 & 2.9 & 78 & 2.1 & 118.0 & MAGIC J2001+435 \\
 &  &  &  &  & 2009-11-09 & $20^{\rmn{h}} 01^{\rmn{m}} 11^{\rmn{s}}$ & $+43\degr 52\arcmin 07\arcsec$ & 0.7 & 51.0 & 2.9 & 54 &  &  &  \\
 &  &  &  &  & 2010-01-10 & $20^{\rmn{h}} 01^{\rmn{m}} 11^{\rmn{s}}$ & $+43\degr 52\arcmin 07\arcsec$ & 0.7 & 57.4 & 2.9 & 51 &  &  &  \\
J2339.6$-$0532 & J2339.7$-$0531 & 81.36 & $-$62.47 & 2.5 & 2009-11-21 & $23^{\rmn{h}} 39^{\rmn{m}} 44^{\rmn{s}}$ & $-05\degr 31\arcmin 13\arcsec$ & 1.9 & 11.9 & 0.8 & 105 & 5.7 & 15.7 &  \\
 &  &  &  &  & 2009-11-21 & $23^{\rmn{h}} 39^{\rmn{m}} 44^{\rmn{s}}$ & $-05\degr 31\arcmin 13\arcsec$ & 1.9 & 21.8 & 0.8 & 77 &  &  &  \\
 &  &  &  &  & 2009-11-25 & $23^{\rmn{h}} 39^{\rmn{m}} 44^{\rmn{s}}$ & $-05\degr 31\arcmin 13\arcsec$ & 1.9 & 46.0 & 0.8 & 53 &  &  &  \\
 &  &  &  &  & 2010-01-12 & $23^{\rmn{h}} 39^{\rmn{m}} 44^{\rmn{s}}$ & $-05\degr 31\arcmin 13\arcsec$ & 1.9 & 56.0 & 0.8 & 48 &  &  &  \\
\end{longtable}

\end{landscape}
\end{scriptsize}
\end{center}

\end{document}